\documentclass{achemso}
\usepackage[version=3]{mhchem}
\usepackage{siunitx}
\usepackage{booktabs}
\usepackage[hidelinks]{hyperref}
\usepackage{amssymb}
\usepackage{bm}
\usepackage{mathtools}
\usepackage[labelfont=bf,labelsep=space]{caption}
\usepackage{multirow}
\DeclareSIUnit\angstrom{\text {Å}}
\DeclareSIUnit[quantity-product = ]\percent{\char`\%}

\author{Ling Zhang}

\affiliation{Department of Chemistry, Molecular Sciences Research Hub, White City Campus, Imperial College London, Wood Lane, London W12 0BZ, UK}
\alsoaffiliation{School of Physics, Beihang University, Beijing, 100191, China}
\alsoaffiliation{Hangzhou International Innovation Institute, Beihang University, Hangzhou 311115, China}

\author{Miao Zhou}
\affiliation{School of Physics, Beihang University, Beijing, 100191, China}
\alsoaffiliation{Hangzhou International Innovation Institute, Beihang University, Hangzhou 311115, China}
\alsoaffiliation{Tianmushan Laboratory, Beihang University, Hangzhou 311115, China}
\email{mzhou@buaa.edu.cn}

\author{Alex M. Ganose}
\affiliation{Department of Chemistry, Molecular Sciences Research Hub, White City Campus, Imperial College London, Wood Lane, London W12 0BZ, UK}
\email{a.ganose@imperial.ac.uk}

\title{Dopability limits in Al-rich AlGaN alloys for far-UVC LEDs}

\begin{document}

\begin{abstract}
Transitioning to solid-state ultraviolet (UV) lighting is critical for reducing global energy utilization to meet net-zero targets. AlGaN-based far-UVC LEDs offer a mercury-free, energy-efficient alternative to conventional mercury lamps, yet their performance is severely bottlenecked by poor carrier injection at Al compositions exceeding 80\%. Point defects are known to significantly affect carrier concentrations and radiative recombination efficiency, however, systematic studies of point defects in AlGaN alloys remain scarce. In this work, we investigate intrinsic and extrinsic defects in high-Al-content Al$_{1-x}$Ga$_x$N alloys ($x$ = 1/6, 1/4, and 1/3). We reveal that explicit alloy modeling and proper treatment of the temperature dependence of the band gap are essential to bring calculated carrier concentrations in line with experimental observations. We uncover that Si dopants preferentially substitute minority Ga atoms, forming compensating negative-\textit{U} \textit{DX} centers in Al-rich environments that severely limit n-type conductivity. We identify carbon as the most detrimental unintentional impurity, while the impact of oxygen and hydrogen is negligible in Si-doped samples typically used for devices. These findings highlight the significance of explicit alloy modeling and provide valuable insights into the design of AlN-based alloys.
\end{abstract}

\section{Introduction}

        Solid-state lighting, spearheaded by light-emitting diodes (LEDs), has fundamentally transformed global energy utilization for household lighting, offering massive reductions in power consumption and carbon footprints compared to traditional incandescent or fluorescent sources.\cite{tsao2010solid} However, realizing comparable energy savings in the ultraviolet (UV) regime remains a significant challenge.\cite{rass2023enhanced} Conventional UV sources, such as mercury vapor lamps, are inherently energy-intensive, bulky, and rely on toxic elements, making them incompatible with sustainable, net-zero objectives. Consequently, developing high-efficiency, mercury-free solid-state UV emitters is imperative for applications spanning environmental sterilization, water purification, and industrial processing.\cite{song2016application,wang2024wafer}

        In this context, aluminum gallium nitride (AlGaN), a ternary alloy in the group III nitride family, has emerged as a key wide-band gap semiconductor material owing to its tunable optical and electronic properties.\cite{Lang2025,Jo2022,Schilling2025,Zhang2005,Fu2019}By varying the Al composition, its direct band gap can be engineered from 3.4 eV (GaN) to 6.2 eV (AlN), enabling emission wavelengths spanning 210–320 nm range and making it well suited for deep-ultraviolet optoelectronic devices.\cite{Jo2022,Zhang2005,Shatalov2012,Hirayama2014,Kneissl2019} In particular, far-UVC radiation (typically 200–235 nm) has recently attracted considerable attention because it can efficiently inactivate microorganisms while causing minimal damage to human skin and eyes\cite{Nozomi2020,Gorlitz2024}. In this regard, Al-rich AlGaN is particularly advantageous for far-UVC emission ($\sim$222 nm), as growth on AlN substrates significantly reduces lattice mismatch and dislocation density, thereby improving crystalline quality and device performance.\cite{Dalmau2011}  

        However, AlGaN-based far-UVC LEDs still exhibit a much lower wall plug efficiency than their near-UV or visible counterparts, primarily due to their limited external quantum efficiency (EQE). \cite{Lang2025,Jo2022} For devices emitting around 222 nm, reported EQEs are typically below 1\%, with some as low as 0.1\% \cite{Guttmann_2019}. Such low efficiency severely limits their practical deployment, particularly in high-intensity disinfection or industrial scenarios. One of the major challenges in improving EQE lies in achieving low-resistivity n- and p-type doping in Al-rich AlGaN layers. \cite{Yang2021,Wang2024} However, for alloys with Al content exceeding 80\%, even n-type doping becomes increasingly difficult due to high dopant activation energies and the potential formation of \textit{DX} centers associated with commonly used dopants such as silicon.\cite{Wang2024} Moreover, the formation of defect complexes between these dopants and intrinsic or unintentional impurities further complicates the doping process and its impact on carrier transport and recombination. \cite{Kataoka2024,Yang2021,Bryan2018,Sarkar2018} Therefore, a detailed understanding of the nature and behavior of these active defects is essential for developing strategies to suppress their detrimental effects.

        While point defects in AlN and GaN have been extensively investigated, \cite{Alden2018,Maki2011,Stampfl2002, zhu2024_AlN,lyons2021first,Diallo2016} systematic studies of defects in AlGaN alloys are relatively scarce. Many previous studies rely on interpolations between the binary end members, rather than direct modeling of the alloy system, \cite{Mirrielees2021,Gordon2014} which may overlook alloy-specific effects such as configurational entropy \cite{tolborg2022}, alloy disorder \cite{baranovskii2022} and short range order \cite{zhang2020,prozheev2025} on defect modeling. Therefore, a comprehensive and quantitative understanding of the defect landscape and its complex donor/acceptor behavior is critical for the further development and optimization of AlGaN-based optoelectronic devices. 

        In this work, we investigate intrinsic and extrinsic point defects in AlGaN alloy, using explicit alloy modeling and a global structure searching strategy. Our results reveal that the intrinsic carrier concentration is substantially higher under N-poor than N-rich conditions, yet remains extremely low in undoped samples, underscoring the necessity of extrinsic n-type dopants such as Si. The inclusion of temperature-dependent band-gap corrections enhances the calculated carrier concentration by nearly two orders of magnitude, bringing theory into closer alignment with experimental observations. Moreover, robust treatment of the experimentally accessible chemical potentials reveal the presence of secondary phases that limit dopability. We highlight the role of carbon as the most detrimental unintentional impurity and further demonstrate that the impact of hydrogen and oxygen is minimized by silicon-doping. Our work outlines a reliable computational framework for understanding defect behavior in AlGaN alloys, and offers valuable guidance for experimental research aimed at optimizing device performance.

\section{Results}
\subsection{Crystal and electronic structure}
\begin{figure}[htbp]
    \centering
    \includegraphics[width=0.8\textwidth]{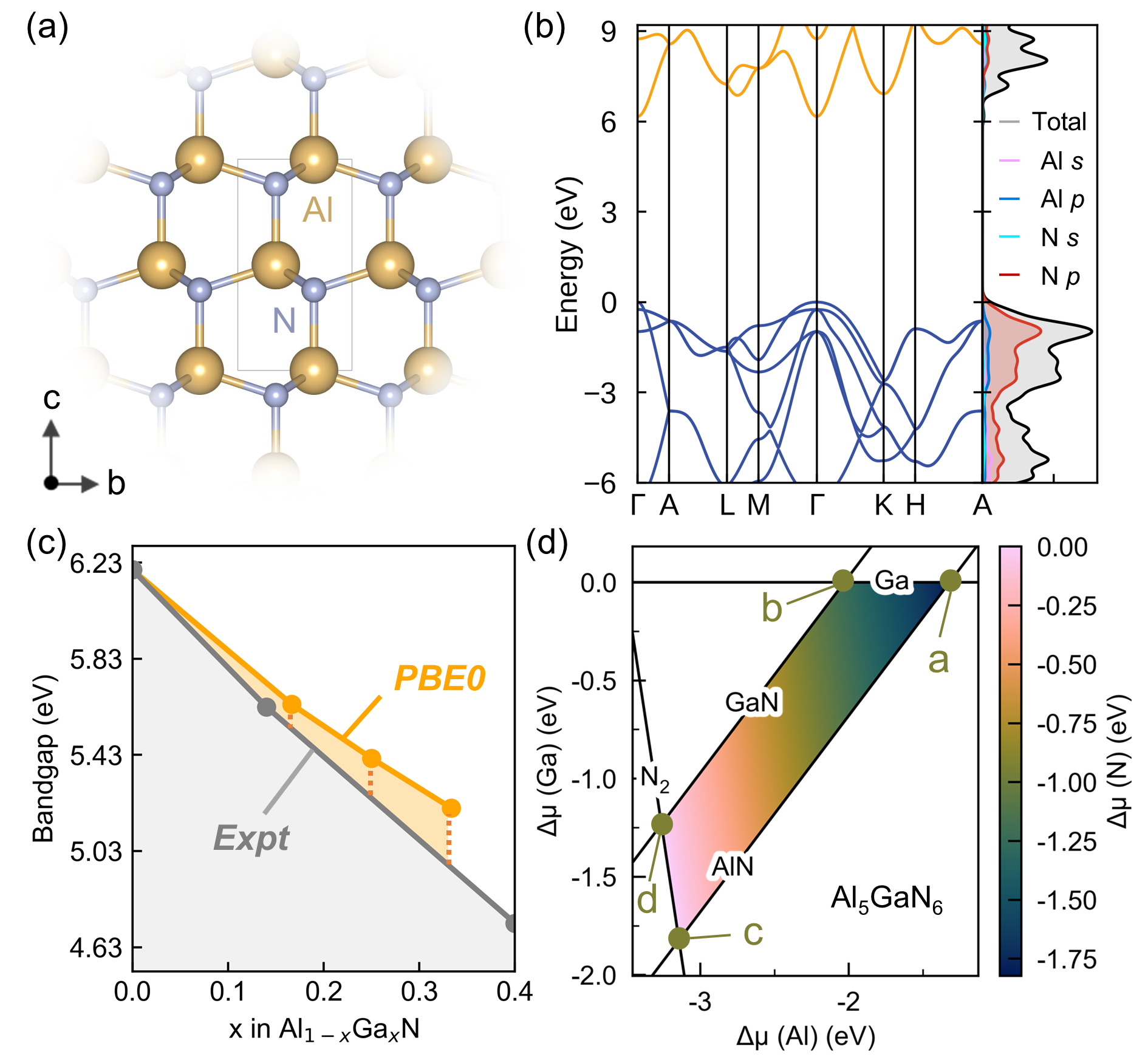}
    \caption{(a) Crystal structure and (b) Electronic band structure alongside the density of states with 0.2 eV Gaussian broadening of wurtzite AlN. (c) The band gap of Al$_{1-x}$Ga$_x$N as a function of $x$, compared with experimental data from ref.\cite{Shan1999}. (d) Al-Ga-N ternary chemical potential diagram showing the stable region of Al$_5$GaN$_6$ due to the limits imposed by competing phases.}
    \label{figure:1}
\end{figure}

        Bulk AlN crystallizes as wurtzite structure at ambient pressure and temperature \cite{Siegel2006}, as shown in Figure~\ref{figure:1}(a). The calculated lattice parameters of AlN are a = b = 3.11 Å and c = 4.97 Å, in close agreement with the experimental values at room temperature (a = b = 3.11 Å, c = 4.98 Å). \cite{Yuri2022} Figure \ref{figure:1}(b) presents the calculated electronic structure of AlN. It is shown that AlN has a direct band gap of 6.20 eV at $\Gamma$ point, in line with previous studies. \cite{Wu2009} The valence band maximum (VBM) is mainly composed of N \textit{p} states, while the conduction band minium (CBM) band shows mixed Al/N character. Figure~\ref{figure:1}(c) shows the calculated and experimental \cite{Shan1999}  band gap of AlGaN with the increase of Ga composition. The band gap and effective masses of Al$_{1-x}$Ga$_x$N as a function of $x$ are presented in Table S1 of Supplementary Section S1. Due to the structural similarity between GaN and AlN,  AlGaN alloys maintain the wurtzite phase upon Ga incorporation. We observe that the band gap of AlGaN decreases progressively as the Ga fraction increases, while the difference between the theoretical and experimental values increases as the Ga fraction increases. This trend originates from the fixed fraction of exact exchange, which is optimized for AlN and is applied uniformly to all alloy compositions in this study.
        
\subsection{Intrinsic point defects}
        Owing to the reduced symmetry of alloy systems, a large number of possible interstitial sites can exist, making a comprehensive sampling of all sites impractical. Therefore, we first investigated all intrinsic point defects in AlN, including \textit{V}\textsubscript{Al}, \textit{V}\textsubscript{N}, Al\textsubscript{N}, N\textsubscript{Al}, Al\textsubscript{$i$} and N\textsubscript{$i$} to exclude defect types with high formation energies. For each defect species, ground-state geometries are identified using the ShakeNBreak \cite{Mosquera2022,Mosquera2023} defect structure searching workflow. The formation energies of all intrinsic defects in AlN under different equilibrium growth conditions are shown in Supplementary Section S2. The formation energies are plotted as a function of the Fermi level \textit{E}\textsubscript{F} , which is referenced to the VBM and ranges to the CBM. The slope of the lines indicate the charge state \textit{q}. We only present the charge state with the minimum formation energy for a given \textit{E}\textsubscript{F}. The charge transition level (CTL, \textit{q}/ \textit{q}') is defined as the Fermi energy at which two charge states \textit{q} and \textit{q'}  have identical formation energies and corresponds to the point where the slope of the formation energy curve changes.  The position of a CTL with respect to the band edges reflects the thermal energy required to ionize the defect. Accordingly, whether a defect behaves as a ``shallow'' or ``deep'' center is governed by how far its CTL lies from the band edges.\cite{squires2026guidelines,gorai2025guide} We find that the formation energies of substitutional and interstitial defects are substantially higher than those of vacancies. For example, at the self-consistent Fermi level under N-poor conditions, the formation energy of N\textsubscript{$i$} and Al\textsubscript{N} are 6.95 eV and 9.09 eV respectively, while \textit{V}\textsubscript{N} is 2.79 eV implying that substitutional and interstitial defects will only be present in low concentrations. Therefore, our subsequent defect analysis on alloy systems focuses exclusively on vacancy defects.

        The Al$_{1-x}$Ga$_x$N alloys are modeled with special quasi-random structures\cite{Zunger1990sqs,Wei1990sqs,Hass1990sqs,VANDEWALLE2013sqs} at different Ga contents ($x$=1/6, 1/4 and 1/3). Figure~\ref{figure:1}(d) and Figure S2 present the Al-Ga-N ternary chemical potential diagram. The phase stability of Al$_{1-x}$Ga$_x$N alloys was determined by calculating the total energy of the competing phases within the Al-Ga-N chemical space using the doped package \cite{Kavanagh2024doped}. Notably, metallic Al does not form a boundary with the Al$_{1-x}$Ga$_x$N phase, implying that the condition $\Delta\mu_{\mathrm{Al}}=0$ is not thermodynamically accessible. This behavior differs from that of pure AlN, in which the Al-rich chemical potential limit is chemically accessible.

\begin{figure}[htbp]
    \centering
    \includegraphics[width=1\textwidth]{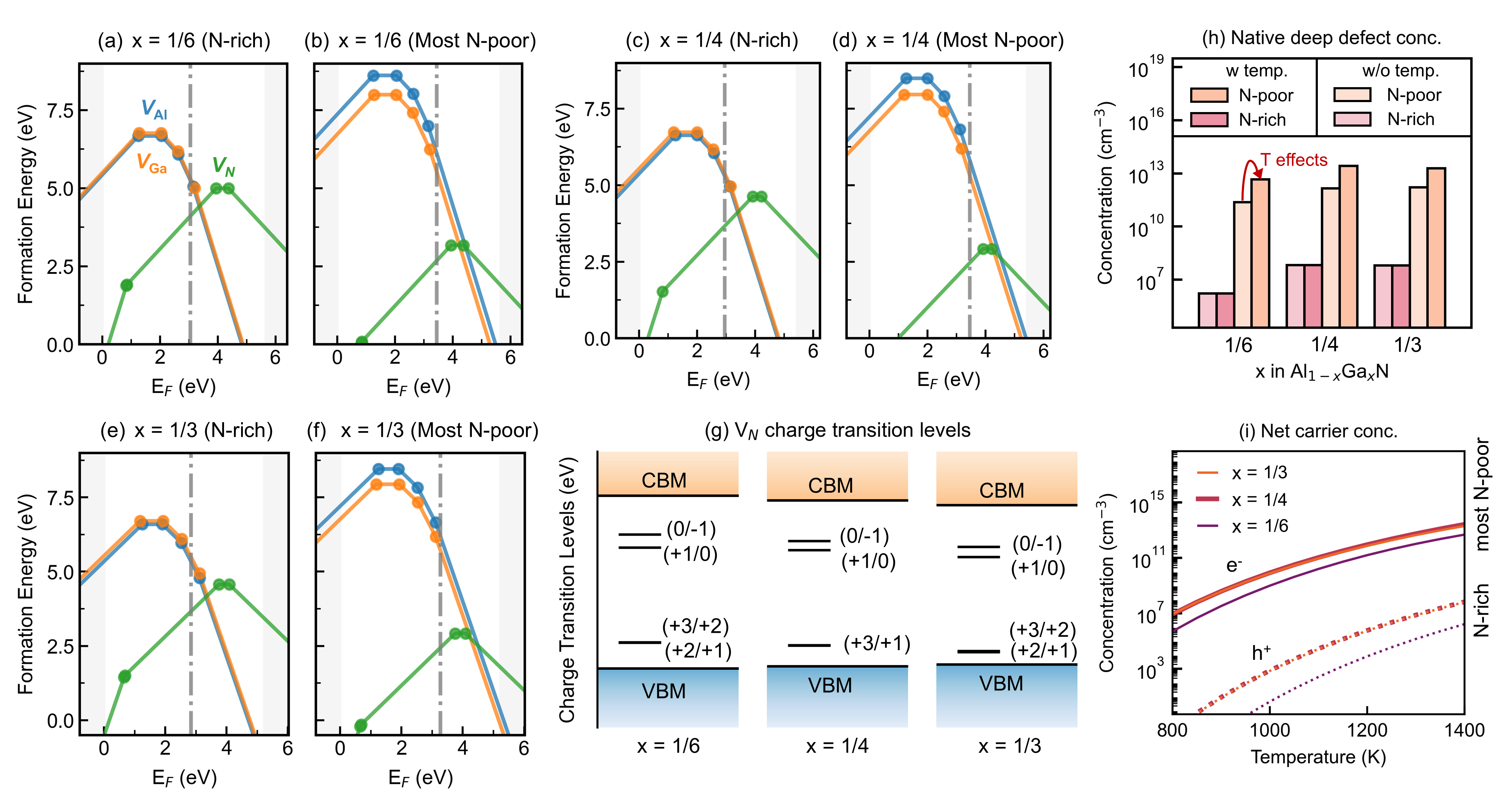}
    \caption{Calculated defect formation energy as a function of the Fermi energy in Al$_{1-x}$Ga$_x$N alloys under N-rich and most N-poor growth conditions for (a,b) $x$ =1/6, (c,d) $x$ =1/4, (e,f) $x$ =1/3. The grey dash lines are the equilibrium Fermi levels, which are determined self-consistently to ensure charge neutrality at T = 1400 K, the typical AlGaN alloy growth temperature. (g) Charge transition levels of nitrogen vacancy in Al$_{1-x}$Ga$_x$N alloys at $x$ = 1/6, 1/4 and 1/3. (h) Total concentration of deep defects in Al$_{1-x}$Ga$_x$N for $x$ = 1/6, 1/4 and 1/3 under N-rich and most N-poor growth conditions with and without the dependence of temperature on band gap. (i) Calculated net carrier concentrations with temperature-dependent band gap renormalization as a function of annealing temperature and quenching at room temperature. The net carrier concentration is $\lvert n_e - n_h \rvert$, where $n_e$ and $n_h$ are electron and hole concentrations, respectively.}
    \label{figure:2}
\end{figure}

        To model point defects in the alloy systems, we evaluated the energies of all symmetry-inequivalent defect sites in the neutral charge state, with the charged defects only calculated in the lowest energy configuration. A detailed description of the defect calculation process is provided in the Methodology Section. The formation energies of intrinsic point defects \textit{V}\textsubscript{Al}, \textit{V}\textsubscript{Ga} and \textit{V}\textsubscript{N} in Al$_{1-x}$Ga$_x$N under both N-rich and N-poor growth conditions are shown in Figure~\ref{figure:2}(a-f). Several general trends are found to persist across different compositions and growth conditions. We find that \textit{V}\textsubscript{Al}, \textit{V}\textsubscript{Ga} and \textit{V}\textsubscript{N} are deep defects with their thermodynamic transition levels far from VBM/CBM. This indicates that they are potential nonradiative recombination centers.  In addition, the defects show amphoteric behavior, switching from donors to acceptors with the change of the Fermi level.  At the equilibrium Fermi energy calculated self-consistently by solving for charge neutrality at 1400 K\cite{VandeWalle2004,Freysoldt2014}, we find that \textit{V}\textsubscript{Al} and \textit{V}\textsubscript{Ga} act as acceptors while \textit{V}\textsubscript{N} behaves as a donor. The formation energy of \textit{V}\textsubscript{N} is substantially lower than that of \textit{V}\textsubscript{Al} and \textit{V}\textsubscript{Ga}, especially under N-poor growth conditions typically employed in experiments. This results in significantly higher defect concentrations (over 12 order of magnitude) for \textit{V}\textsubscript{N} compared to other vacancies.
        
        To understand the behavior of the nitrogen vacancy in undoped Al$_{1-x}$Ga$_x$N, we plot the CTL diagram in Figure~\ref{figure:2}(g). The energy of the CBM (VBM) decreases (increases) as the fraction of Ga ($x$) increases. Meanwhile, the CTLs shift within the band gap, with concurrent impacts on the carrier concentration. For the (0/-1) transition, the decrease in CTL combined with the upward shift of the VBM enables easier hole ionization from the defect state, thereby rendering \textit{V}\textsubscript{N} defect more shallow. For example, for the (+1/0) transition, the energy difference between CBM and the defect states is 1.70, 1.63 and 1.37 eV for $x$=1/6, 1/4 and 1/3, respectively. As noted previously, this trend is responsible for the more facile n-type doping of high-Ga content AlGaN samples\cite{Sarkar2018,Freysoldt2014}.

        Figure~\ref{figure:2}(h) presents the calculated defect concentrations in undoped Al$_{1-x}$Ga$_x$N alloys under both N-rich and N-poor conditions. We apply the frozen-defect approximation to consider the influence of temperature on the defect and carrier concentrations. The frozen-defect approximation simulates a synthetic scenario in which the material is annealed at elevated temperatures and subsequently quenched to room temperature. \cite{Squires2023} Here, defects are assumed to form in equilibrium at the annealing temperature and persist during quenching to room temperature due to substantial kinetic barriers that hinder diffusion and annihilation. Moreover, AlN exhibits significant temperature-dependent band gap renormalization. At the synthesis temperature of \SI{1400}{\kelvin}, the band gap is lowered by approximately \SI{2.51}{\electronvolt}.\cite{Nepal2005} Such variation in the band gap significantly influences the Fermi-level position, thereby modifying the formation energies of charged defects and ultimately altering the predicted defect concentrations  (demonstrated in more detail in Supplementary Section S4).\cite{Irea2023} We therefore explicitly include the temperature dependence of the band gap at annealing conditions in our analysis. As shown in Figure ~\ref{figure:2}(h), without applying temperature-dependent corrections, the deep defect concentrations for all systems are below 10\textsuperscript{9} cm\textsuperscript{-3} under N-rich condition. Such a low concentration is at or below the detection limit of state-of-the-art characterization techniques. When including temperature-dependent band gap renormalization, the deep defect concentrations increase to $\sim$$10^{12}$ to $10^{13}$ cm\textsuperscript{-3}. These results highlight the importance of properly accounting for the temperature-dependence of the band gap when calculating defect concentrations in these systems, which has been neglected in prior studies\cite{Mirrielees2021,Washiyama2021,GE2025}. 

        Achieving a high carrier concentration is crucial for far-UVC LED devices as it not only ensures an adequate electron supply for recombination within the multi-quantum wells but also modulates the Fermi level to facilitate the formation of low-resistance Ohmic contacts. \cite{Lang2025} Figure~\ref{figure:2}(i) presents the net carrier concentration as a function of the annealing temperature with quenching at room temperature. The net carrier concentration is defined as $|n_e-n_h|$, where $n_e$ and $n_h$ represent the concentrations of electrons and holes, respectively. When $n_e > n_h$ ($n_h > n_e$), the corresponding plots are designated as $e^-$ ($h^+$). Under the most N-poor conditions, the alloys exhibit n-type behavior, yielding net electron concentrations of between $10^{12}$ to $10^{13}$ cm$^{-3}$, in line with the defect concentrations discussed above. Conversely, N-rich growth conditions results in slight p-type doping but with net carrier concentrations over 4 orders of magnitude lower. These results are consistent with experimental observations, where measurable carrier densities become difficult to achieve once the Al content exceeds ~0.6. \cite{Nam2002,Polyakov1998} Such low  intrinsic carrier concentrations in undoped Al$_{1-x}$Ga$_x$N are insufficient for high power LED devices (typically requiring at least 10$^{18}$ cm$^{-3}$) \cite{Lang2025}. Consequently, the introduction of extrinsic dopants is essential to enhance the performance of AlGaN alloys.

\subsection{Extrinsic Si doping defects}

\begin{figure}[htbp]
    \centering
    \includegraphics[width=1\textwidth]{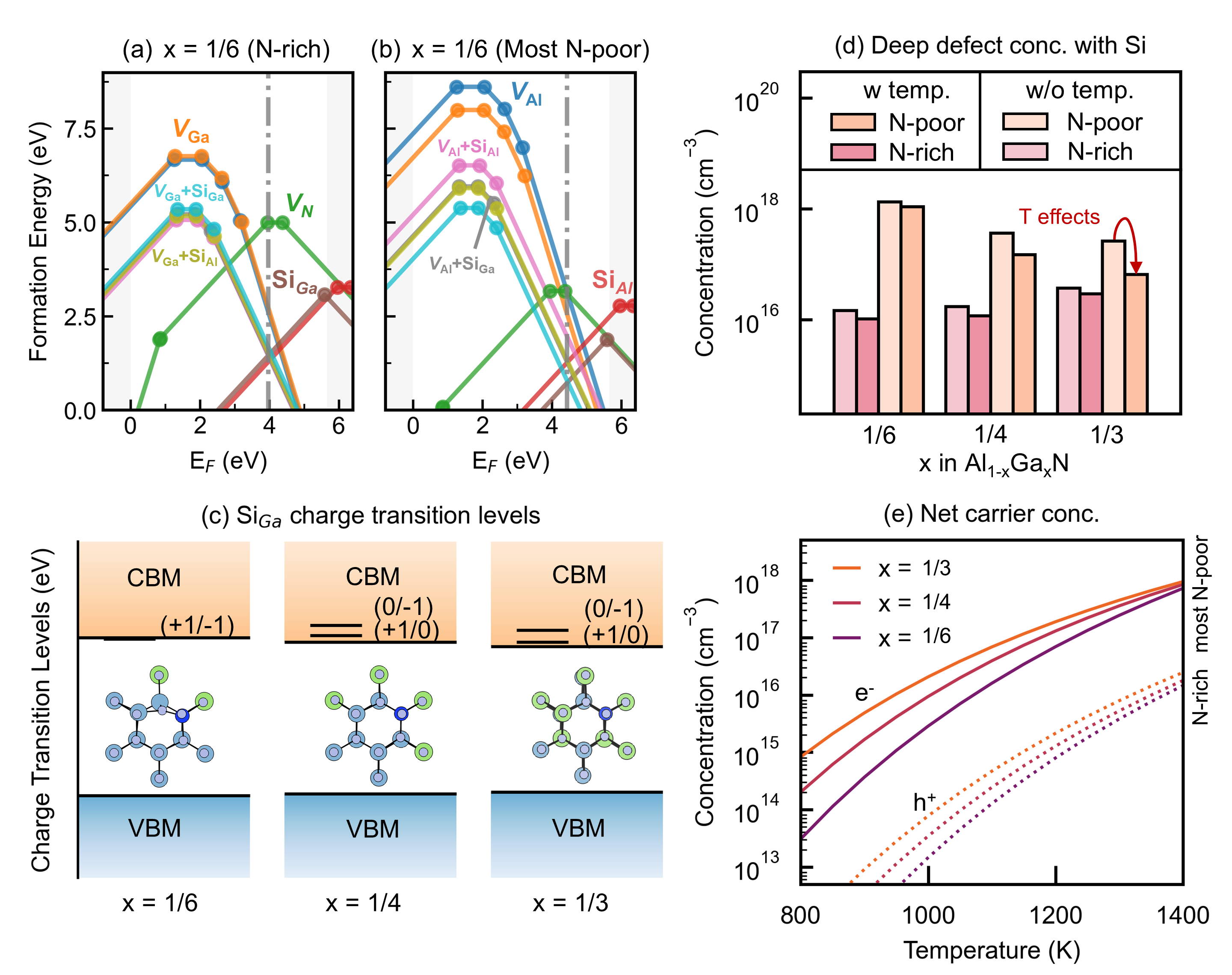}
    \caption{Calculated defect formation energy as a function of the Fermi energy in Si-doped Al$_5$GaN$_6$ alloys under (a) N-rich and (b) most N-poor growth conditions. The equilibrium Fermi energy is calculated at an annealing temperature of 1400 K. (c) Charge transition levels and local structures of Si\textsubscript{Ga} in Al$_{1-x}$Ga$_x$N alloys at $x$ = 1/6, 1/4 and 1/3. The light blue, dark blue, green and grey spheres represent Al, Si, Ga and O atoms, respectively. (d) Total concentration of deep defects (\textit{V}\textsubscript{Al}, \textit{V}\textsubscript{Ga}, \textit{V}\textsubscript{N}, \textit{V}\textsubscript{Al}+Si\textsubscript{Al}, \textit{V}\textsubscript{Al}+Si\textsubscript{Ga}, \textit{V}\textsubscript{Ga}+Si\textsubscript{Al} and \textit{V}\textsubscript{Ga}+Si\textsubscript{Ga}) in Al$_{1-x}$Ga$_x$N for $x$ = 1/6, 1/4 and 1/3 under N-rich and most N-poor growth conditions with and without the influence of temperature on band gap. (e) Calculated net carrier concentrations on the band gap as a function of annealing temperature and quenching at room temperature. The temperature-dependent band gap renormalization is included. The net carrier concentration is $\lvert n_e - n_h \rvert$, where \textit{$n_e$} and \textit{$n_h$} are electron and hole concentrations, respectively.}
    \label{figure:3}
\end{figure}

        Previous experimental and computational studies have highlighted the role of n-type Si doping as an effective route to introduce free carriers in AlGaN\cite{Wang2024,Bryan2018,Harris2018,vandewalle1999}. Figure~\ref{figure:3}(a) and 3(b) present the calculated formation energies of intrinsic vacancies, Si substitutions (Si\textsubscript{Al} and Si\textsubscript{Ga}), and defect complexes (\textit{V}\textsubscript{Al}+Si\textsubscript{Al}, \textit{V}\textsubscript{Al}+Si\textsubscript{Ga}, \textit{V}\textsubscript{Ga}+Si\textsubscript{Al} and \textit{V}\textsubscript{Ga}+Si\textsubscript{Ga}) in Si-doped Al$_5$GaN$_6$. Corresponding defect formation energies of Al$_2$GaN$_3$ and Al$_3$GaN$_4$ are shown in Figure S4 of Supplementary Section S5. For substitutional defects, the formation energies of Si\textsubscript{Al} and Si\textsubscript{Ga} are 1.26 and 0.72 eV under N-poor conditions at the equilibrium Fermi level, indicating that Si substitutional defects are readily formed during synthesis. In the system with the highest Al content, Al\textsubscript{5}GaN$_6$, the Si\textsubscript{Ga} exhibits a two-electron negative-\textit{U} transition situated approximately 65 meV below the CBM, in excellent agreement with the experimental activation energy of 67 meV for Al\textsubscript{0.82}Ga\textsubscript{0.18}N \cite{Mehnke2016}. Structural analysis reveals the formation of a Si\textsubscript{Ga} \textit{DX} center --- a substitutional defect in which the dopant atom is displaced from its lattice site while trapping extra electrons \cite{Gordon2014,Joshua2019} --- with the nearest neighbor N site moving close to the Si atom [Figure~\ref{figure:3}(c)]. This \textit{DX} center has previously been associated with the pronounced reduction of carrier concentration in Al-rich AlGaN alloys.\cite{Trinh2014,Mehnke2016} Interestingly, even though Al is the majority metal in Al-rich alloys, Si prefers to substitute on the Ga site which alone is responsible for the \textit{DX} structural distortion that limits n-type conductivity. Upon moving to greater Ga content, the negative-\textit{U} behavior is lost, and the (0/-1) and (+1/0) CTLs shift inside the conduction band for $x$=1/4 and 1/3, with no obvious structural distortion. In contrast, the CTL of Si\textsubscript{Al} is resonant in the conduction band for all three compositions as demonstrated in Figure S5 of the Supplementary Information. Careful examination of the local structure reveals that there are 5 (2), 5 (4), 7 (7) Ga atoms among the next-nearest neighboring twelve metal atoms of the doped Si atom for Si\textsubscript{Al} (Si\textsubscript{Ga}) in Al\textsubscript{5}GaN\textsubscript{6}, Al\textsubscript{3}GaN\textsubscript{4} and Al\textsubscript{2}GaN\textsubscript{3}, respectively. This suggests that Ga-rich environments stabilize the donor character of Si, whereas Al-rich environments favor the formation of acceptor-like \textit{DX} centers, as recently proposed by Prozheev \textit{et al.} \cite{prozheev2025}. For the defect complexes, all show deep transition levels with amphoteric character. The formation energies of the complexes are lower than single metal vacancies across all Fermi levels, indicating that the combination of metal vacancies with Si substitutional defects is highly favorable.

         Si doping shifts the equilibrium Fermi level closer to the CBM under both N-rich and most N-poor conditions, leading to higher defect and carrier concentrations. Figure~\ref{figure:3}(d) presents the calculated deep defect concentrations in Si-doped Al$_{1-x}$Ga$_x$N. Below, we focus on N-poor conditions since these are most relevant in experiments, while the analysis of the N-rich conditions is provided in the Supplementary Section S5.  Without applying temperature-dependent band gap renormalization, the deep defect concentration for all system are  $\sim$$10^{17}$ to $10^{18}$ cm\textsuperscript{-3}. By including temperature-dependent band gap renormalization, the deep defect concentrations decrease by 1.2 times for $x$=1/6, 2.43 times for $x$=1/4, and 40 times for x=1/3. The reduction of these deep defect concentrations facilitates an increase in carrier density.
        Figure~\ref{figure:3}(e) and Figure S6 show the net carrier concentration in Si-doped Al$_{1-x}$Ga$_x$N alloys with and without temperature-dependent corrections, respectively. The alloys are predicted to exhibit n-type conductivity under N-poor conditions and p-type conductivity under N-rich conditions. The net carrier concentrations are close to $10^{18}$ cm$^{-3}$ at an annealing temperature of 1400 K for Si-doped Al$_{1-x}$Ga$_x$N  under N-poor conditions. This is in reasonable agreement with experimental measurements ($\sim$10$^{19}$ cm$^{-3}$). \cite{Wang2024} We note that, without inclusion of temperature band-gap renormalization (as assumed in previous defect studies), the concentrations drop by 2-3 orders of magnitude, yielding a significant discrepancy with experiment. An analogous phenomenon has recently been observed in CdTe and Al-doped SrTiO$_3$.\cite{Kavanagh2024,Ogawa2025}

\subsection{Unintentional extrinsic defects}

\begin{figure}[htbp]
    \centering
    \includegraphics[width=1\textwidth]{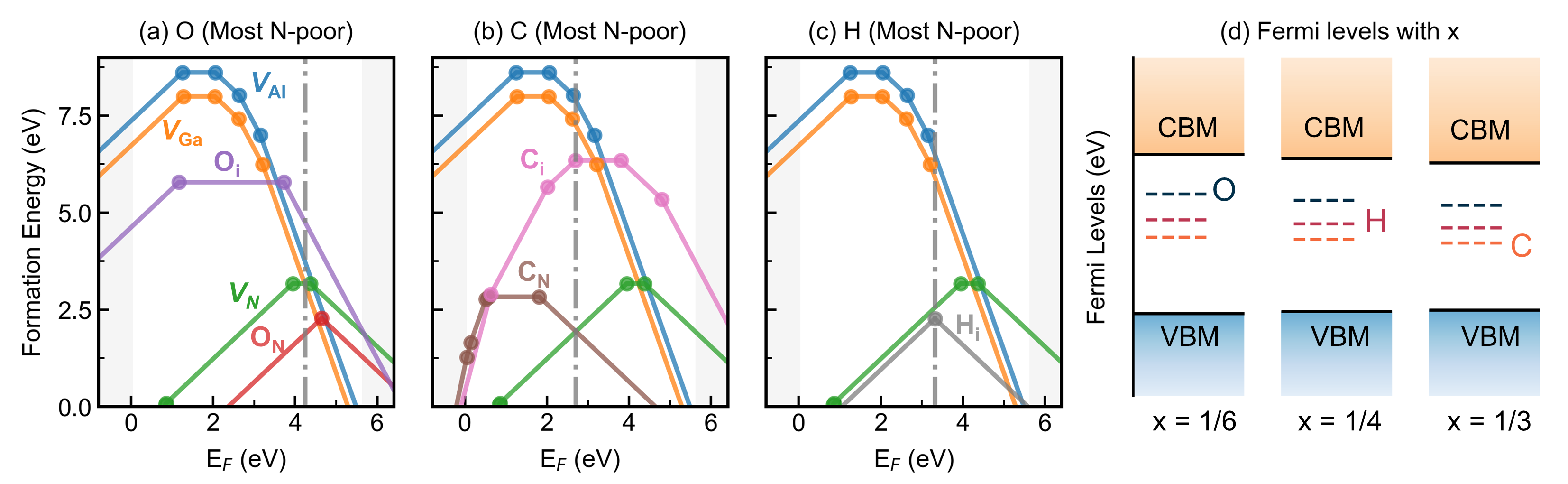}
    \caption{Formation energies of unintentional extrinsic point defects (a) O, (b) C and (c) H in Al$_5$GaN$_6$ alloys under most N-poor growth conditions. The equilibrium Fermi energy is calculated at an annealing temperature of 1400 K. (d) The equilibrium Fermi levels of O, H, C doped Al$_{1-x}$Ga$_x$N alloys at $x$ = 1/6, 1/4 and 1/3.}
    \label{figure:4}
\end{figure}

        Unintentional incorporation of light impurities such as oxygen, carbon, and hydrogen is frequently observed in AlGaN alloys, and is known to strongly influence their electronic and optical performance. \cite{Van1998,Joshua2019,Yan2025} \cite{Kojima2020,zhao2025} We present the defect formation energies of interstitial (O\textsubscript{$i$}, C\textsubscript{$i$} , H\textsubscript{$i$}) and substitutional defects (C\textsubscript{N} and O\textsubscript{N}) in Al\textsubscript{5}GaN\textsubscript{6} for N-poor conditions in Figure~\ref{figure:4}(a–c). We do not consider H\textsubscript{N} due to the large difference in atomic radii between H and N. All defects introduce deep levels in the band gap, and can potentially act as non-radiative recombination centers. Each impurity species shows amphoteric behavior, having stable positive and negative charge states. O\textsubscript{$i$} exhibits high formation energies ($>$ 2.65 eV for all possible Fermi levels), suggesting that it will not form in high concentrations and the influence on carrier concentration will be limited. In contrast, the formation energy of O\textsubscript{N} is relatively low ($<$ 2.5 eV), with a deep negative-\textit{U} transition level 1 eV below the conduction band edge. For C doped Al\textsubscript{5}GaN\textsubscript{6}, similar to the case in O doped Al\textsubscript{5}GaN\textsubscript{6}, the formation energy of C\textsubscript{$i$} is high whereas that of C\textsubscript{N} is relatively low. However, unlike the donor-like O\textsubscript{N}, C\textsubscript{N} introduces acceptor that shifts the equilibrium Fermi level away from the CBM, even with explicit Si-doping, thereby exerting a strong detrimental effect on carrier concentration. For H\textsubscript{$i$}, the formation energy is relatively low, and presents negative-\textit{U} behavior involving a two-electron process. Meanwhile, H\textsubscript{$i$} behaves as effective compensating centers, since it lowers the Fermi level significantly toward mid-gap. Notably, the impact of H$_i$ is dependent on co-doping, while it has great effect in the intrinsic defect system, its presence in Si-doped samples shows little effects. This is because its contribution is negligible compared to the much stronger compensating role of Si-related complexes. 

We further investigate the equilibrium Fermi level within the band gap in O, H, and C doped Al$_{1-x}$Ga$_x$N alloys, as shown in Figure~\ref{figure:4}(d). With increasing Ga composition from 1/6 to 1/3, the band gap narrows, resulting from an upward shift of the VBM and a downward shift of the CBM. For O-doped systems, the energy gap between the Fermi level and CBM increases from 1.41 to 1.51 eV with the increase of Ga composition, suggesting reduced donor efficiency at higher Ga contents. Conversely, in C-doped systems, the energy gap between the Fermi level and CBM decreases from 2.95 to 2.84 eV with the increase of Ga composition, implying the stronger compensating effect with increased Ga fraction. For H-doped systems, the Fermi-level position is insensitive to $x$, with the energy gap between CBM and Fermi level decreasing slightly with the increase of Ga fraction, however as previously noted its impact with explicit Si co-doping is minimal. These results indicate that among unintentional defects, C\textsubscript{N} defects act as significant compensating centers, strongly influencing the carrier concentration and therefore should be minimized during material growth.

\section{Conclusion}
        In summary, we employed alloy modeling and first-principles defect calculations to systematically investigate both intrinsic and extrinsic point defects in Al-rich AlGaN alloys. Si substitutional defects are identified as the primary shallow donors, whereas complex defects act as dominant compensating centers that critically limit electron concentrations. Among unintentional impurities, carbon incorporation is shown to be particularly detrimental. C$_\mathrm{N}$ induces a Fermi-level shift exceeding 1 eV, drastically suppressing carrier density and thereby undermining the efficiency of AlGaN-based light-emitting devices. More broadly, we demonstrate that traditional defect calculations relying on interpolations of binary end-members and 0 K band gaps are wholly insufficient for accurate predictions in wide-band gap solid solutions. We reveal that explicit alloy modeling and proper treatment of the temperature dependence of the band gap are essential for quantitatively reliable predictions of carrier statistics, factors overlooked in most previous studies. These findings not only clarify the complex role of intrinsic and extrinsic defects in AlGaN alloys, but also provide new guidance for both theoretical modeling and experimental strategies in III-nitride optoelectronics.
        
\section{Methodology}
\subsection{First-principles calculations}
        All first-principles calculations were performed within the framework of Kohn-Sham density functional theory (DFT) \cite{Kohn1965,Dreizler1990} using the Vienna Ab Initio Simulation Package (VASP). \cite{Kresse1996vasp,Kresse1999}  The range-separated screened hybrid DFT functional of PBE0 \cite{adamo1999pbe0} was used for all structural relaxation and static calculation, with 23\% exact exchange fraction to reproduce the experimental band gaps and lattice parameters of AlN (Table S5). The projector augmented-wave (PAW) \cite{blochl1994projector} method was employed with the cutoff of 400 eV. All structures were fully relaxed until the residual forces were less than 0.01 eV / \AA. To avoid Pulay stress, the energy cutoff was increased to 520 eV. The $2 \times 2 \times 2$ $\Gamma$-centered \textit{k}-point mesh was used for all geometry optimization, while a dense $4 \times 4 \times 4$ $\Gamma$-centered \textit{k}-point mesh was used for static calculations to ensure a well-converged density of states. Charge carrier effective masses were obtained from non-parabolic fitting of the electronic band edges using \textit{amset} \cite{Ganose2021} and electronic band structure diagrams were generated using sumo \cite{Ganose2018sumo}.

\subsection{Alloy modeling}
        We used a 96 atom supercell by expanding the primitive unit cell of AlN by a matrix of [[4,2,0],[2,4,0],[-2,-2,2]], which ensures the minimum distance between periodic images of 10~\AA. The Al$_{1-x}$Ga$_x$N alloy was modeled with special quasi-random structures (SQSs) \cite{Zunger1990sqs,Wei1990sqs,Hass1990sqs,VANDEWALLE2013sqs} in icet tool \cite{Angqvist2019icet,Ekborg-Tanner2024}. The SQS is generated through a stochastic exploration of supercell configurations, aiming to reproduce the pairwise correlations of an ideal random alloy. The optimal SQS is identified by minimizing the root-mean-square deviation from the target random correlations. Three 96-atom SQS supercells were constructued to model the Al$_{1-x}$Ga$_x$N alloys, with a range of up to 8.0 and 7.0 \AA{} for pair and triplet clusters, respectively. 

Since the alloy is unstable at 0 K, the calculated energy from DFT can not be directly used to compute the elemental chemical potentials without entropy correction. Due to the usually small contribution and high computational expenses of vibrational contributions, we only considered the configurational entropy contribution to $\Delta{\textit{S}}$, which is,
\begin{equation}
\Delta \mathrm{\textit{S}} = -\textit{k}_B[xln(x)+(1-x)ln(1-x)].
\end{equation}

The mixing free energy was then calculated using 
$\Delta{\textit{F}}$ = $\Delta{\textit{H}_m}$ - T$\Delta{\textit{S}}$, where $\Delta{\textit{H}_m}$ is the alloy mixing enthalpy calculated by the following equation \cite{sinnott2019}:
\begin{equation}
\Delta \mathrm{\textit{H}_m} = \mathrm{\textit{H}_{alloy}} - x\mathrm{\textit{H}_{GaN}} - (1-x)\mathrm{\textit{H}_{AlN}},
\end{equation}
where \textit{H}\textsubscript{alloy} is the enthalpy of the alloy, \textit{H}\textsubscript{AlN} and \textit{H}\textsubscript{GaN} are the enthalpies of AlN and GaN, respectively. 

\subsection{Defect modeling}
        We used doped Python package \cite{Kavanagh2024doped} for input file generation and parsing, analysis and plotting of defect calculations. The ShakeNbreak \cite{Mosquera2023,Mosquera2022,mosquera-lois_search_2021} defect structure-searching method was employed in all cases with local bond distortions of both compression and stretching between 0\% and 60\% with 10\% as an interval. It has been proved to efficiently map complex defect potential energy surfaces and to identify ground-state structures. The charge carrier concentrations under thermodynamic equilibrium are calculated using the total charge neutrality condition.

        Due to the \textit{P}1 symmetry of the SQS, in which each atom occupies a unique crystallographic environment, the 96-atom supercell contains 96 distinct Wyckoff sites. For each symmetry-inequivalent site, only the neutral charge state was initially evaluated to determine the most energetically favorable defect configurations. For those sites with the lowest formation energies in the neutral charge state, a comprehensive set of defect calculations was carried out, considering all possible charge states to identify accurate thermodynamic transition levels. Within the periodic supercell framework, the formation energy $\Delta E_{\mathrm{D},q}$ of defect $D$  in charge state $q$ is computed according to the equation,
\[
\Delta E_{\mathrm{D},q} = E_{\mathrm{D},q} - E_{\mathrm{host}} + \sum_i n_i \mu_i + q E_\mathrm{F} + E_{\mathrm{corr}}
\]
where $E_{\mathrm{D},q}$ and $E_{\mathrm{host}}$ denote the total energies of the supercell with and without defect, respectively. $E_\mathrm{F}$ is the Fermi energy referenced to the VBM. $n_i$ represents the number of atoms of element i added ($n_i < 0$) or removed ($n_i > 0$) from the supercell to form defect D. The last term $E_{\mathrm{corr}}$ represents a post-processing correction introduced to treat the spurious interactions originating from the use of a finite-size supercell. In this work, the Kumagai-Oba (eFNV) \cite{Kumagai2014} charge correction method was employed, as implemented in the doped package \cite{Kavanagh2024doped} through the pydefect \cite{Kumagai2021} API. The corresponding total dielectric tensor was listed in Table S6.

        Equilibrium concentrations of defect and carriers were determined by thermodynamic analyses and constrained by charge neutrality\cite{VandeWalle2004,Freysoldt2014}. Notably, since the band gap of AlGaN is dependent on the temperature, the relative energies of defects and carriers changes accordingly. These changes in the band edge positions alter the defect and carrier formation energies and their equilibrium concentrations, leading to a corresponding shift in the Fermi level. Therefore, the `frozen defect approximation' employed in the doped package\cite{Kavanagh2024doped} was adopted to determine the defect and charge-carrier concentrations, and the corresponding self-consistent Fermi level. The total defect concentrations at the given synthesis/annealing temperature were calculated and subsequently kept constant. At room temperature, only the distribution among different charge states of a given defect is allowed to re-equilibrate under the constraint of the fixed total defect concentration. The band gap values were obtained by extrapolating experimental measured band gaps reported in literature using experimentally determined, composition-dependent Varshni parameters\cite{Nepal2005}.

\section{Acknowledgments}

We thank Se\'{a}n Kavanagh for helpful discussions.
A.M.G.~was supported by EPSRC Fellowship EP/T033231/1.  L. Z. and M. Z. acknowledges the Fundamental Research Funds for Central Universities, the Natural Science Foundation of Zhejiang Province (Grant Nos. LZ22A040004), the National Key R\&D Program of China (Grant No. 2022YFF0708800), the National Natural Science Foundation of China (Grant Nos. 11674042). L. Z. acknowledges the support of the international joint doctoral education fund of Beihang University.
We acknowledge computational resources and support provided by the Imperial College Research Computing Service (\url{http://doi.org/10.14469/hpc/2232}).

\bibliography{refs}

\end{document}


\renewcommand{\thefigure}{S\arabic{figure}}
\renewcommand{\thetable}{S\arabic{table}}

\section{S1 Band gap and effective masses of Al$_{1-x}$Ga$_x$N}
\begin{table}[h!]
\centering
\renewcommand{\arraystretch}{1.2}
\begin{tabular}{cccccc}
\toprule
\multirow{2}{*}{$x$} & \multirow{2}{*}{bandgap (eV)} 
& \multicolumn{2}{c}{$m_e$ ($m_0$)} 
& \multicolumn{2}{c}{$m_h$ ($m_0$)} \\
\cmidrule{3-6}
& & $x,y$ & $z$ & $x,y$ & $z$ \\
\midrule
0 (AlN)   & 6.20 & 0.361 & 0.255 & 2.599 & 0.254 \\
1/6& 5.64 & 0.282 & 0.255 & 2.973 & 0.305 \\
1/4& 5.42 & 0.270 & 0.253 & 3.117 & 0.350 \\
1/3& 5.21 & 0.262 & 0.240 & 2.854 & 0.363 \\
1 (GaN)   & 3.40 & 0.255 & 0.172 & 1.661 & 0.444 \\
\bottomrule
\end{tabular}
\caption{Band gap and effective masses ($m_e$, $m_h$) of Al$_{1-x}$Ga$_x$N as a function of $x$.}
\end{table}

\clearpage
\section{S2 AlN defect formation energies}
\begin{figure}[htbp]
    \centering
    \includegraphics[width=0.8\textwidth]{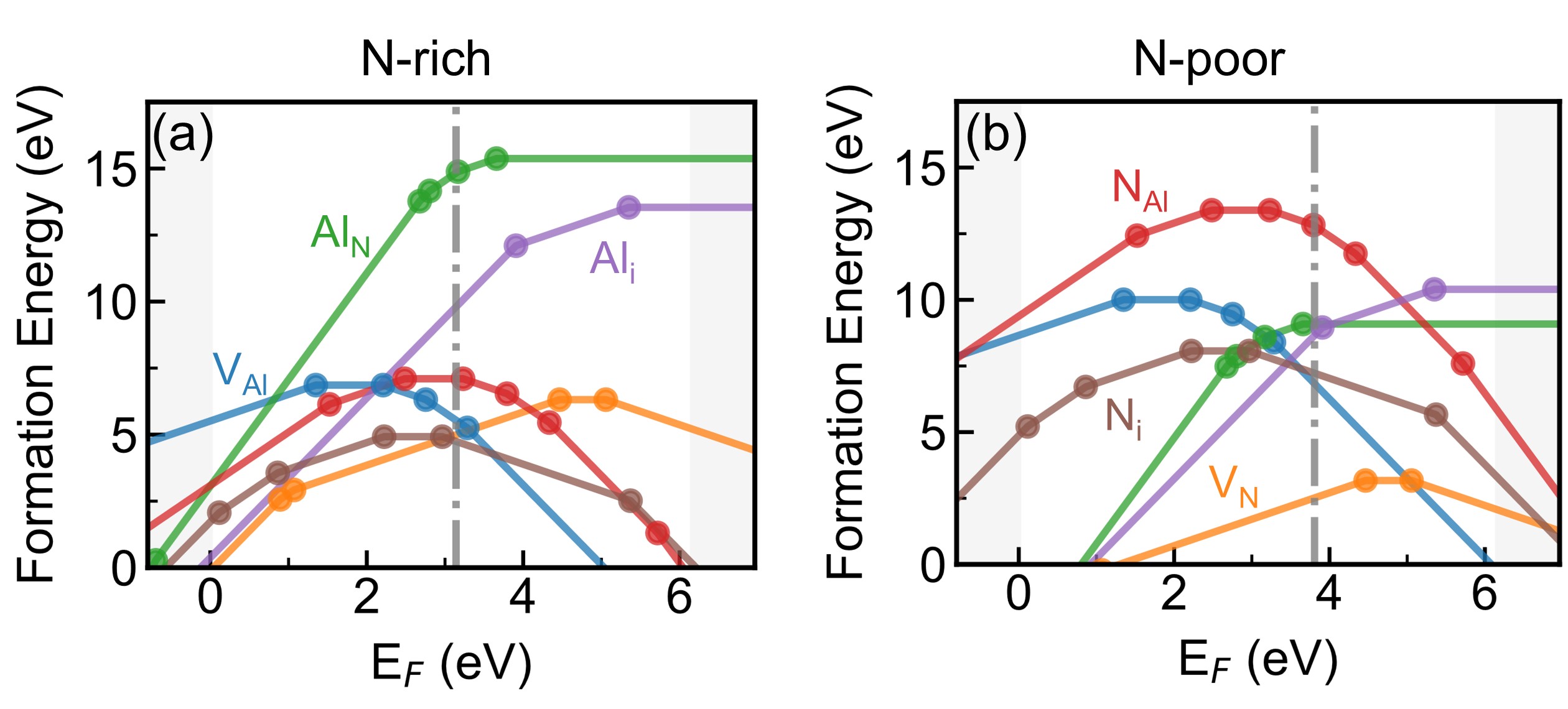}
    \caption{Calculated defect formation energy as a function of the Fermi energy in AlN under (a) N-rich and (b) N-poor conditions.}
    \label{figure:S1}
\end{figure}

\clearpage
\section{S3  Three phase equilibrium regions of Al$_{1-x}$Ga$_x$N}
        The energies of all possible competing phases were computed to determine the stable region of the elemental chemical potentials, where each $\Delta$$\mu_i$ represents the chemical potential of each species. The sum of $\Delta\mu_{\mathrm{Al}}$, $\Delta\mu_{\mathrm{Ga}}$ and $\Delta\mu_{\mathrm{N}}$ is given by the formation energy of Al$_{1-x}$Ga$_x$N as, 
\begin{equation}
(1-x)\Delta \mu_{\mathrm{Al}} + x\Delta \mu_{\mathrm{Ga}} + \Delta \mu_{\mathrm{N}} = \Delta E_{f}(\mathrm{Al_{1-\textit{x}}Ga_\textit{x} N})
\end{equation}
        In addition, $\Delta\mu_{\mathrm{Al}}$, $\Delta\mu_{\mathrm{Ga}}$ and $\Delta\mu_{\mathrm{N}}$ should not stabilize other competing phases,
\begin{equation}
\Delta \mu_{\mathrm{Al}} + \Delta \mu_{\mathrm{N}} \le \Delta E_{f}(\mathrm{AlN})
\end{equation}
\begin{equation}
\Delta \mu_{\mathrm{Ga}} + \Delta \mu_{\mathrm{N}} \le \Delta E_{f}(\mathrm{GaN})
\end{equation}
\begin{equation}
\Delta \mu_{\mathrm{Al}} \le \Delta E_{f}(\mathrm{Al}), \Delta \mu_{\mathrm{Ga}} \le \Delta E_{f}(\mathrm{Ga}), 2\Delta \mu_{\mathrm{N}} \le \Delta E_{f}(\mathrm{N_2})
\end{equation}
        These thermodynamic constraints define the ranges of the chemical potentials $\Delta\mu_{\mathrm{Al}}$, $\Delta\mu_{\mathrm{Ga}}$ and $\Delta\mu_{\mathrm{N}}$ under which Al$_{1-x}$Ga$_x$N remains stable. As shown in Figure 1(d) and Figure S2, this stability region is bounded by four limit phases,\textit{ i.e.}, Ga, AlN, GaN and N\textsubscript{2} and delineated by the four limit conditions (a-d). Tables S2--S4 show the chemical potentials for Al, Ga, and N. Al$_{1-x}$Ga$_x$N is found to be thermodynamically stable across an narrow region of the Al-Ga-N chemical potential space, with $\Delta\mu_{\mathrm{Al}}$, $\Delta\mu_{\mathrm{Ga}}$ and $\Delta\mu_{\mathrm{N}}$ spanning approximately only 2 eV.
\begin{figure}[htbp]
    \centering
    \includegraphics[width=0.8\textwidth]{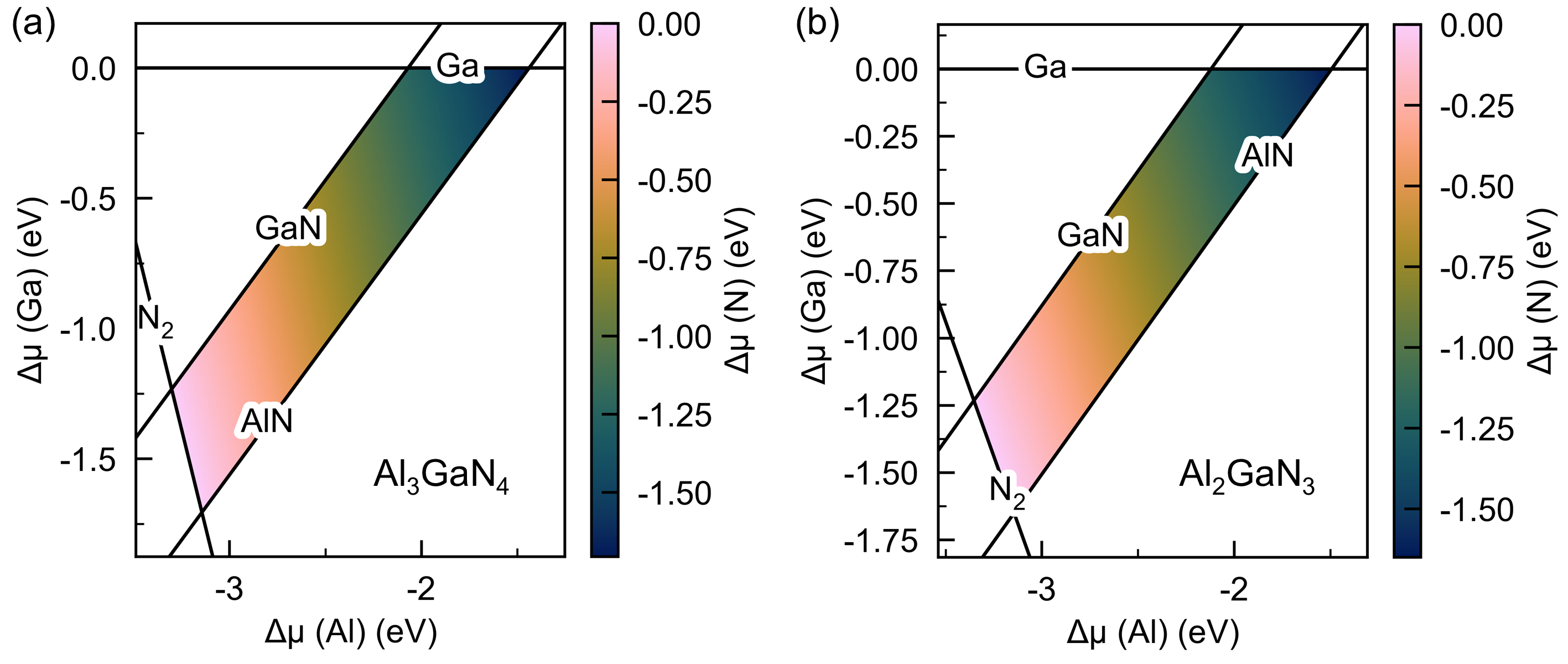}
    \caption{The Al-Ga-N ternary chemical potential diagram showing the stable region of (a) Al$_3$GaN$_4$ and (b) Al$_2$GaN$_3$ due to the limits imposed by competing phases.}
    \label{figure:S2}
\end{figure}
\begin{table}[h!]
\centering
\renewcommand{\arraystretch}{1.2}
\begin{tabular}{cccc}
\toprule
Limit & $\Delta\mu_{\mathrm{Al}}$ & $\Delta\mu_{\mathrm{Ga}}$ & $\Delta\mu_{\mathrm{N}}$ \\
\midrule
Al\textsubscript{5}GaN\textsubscript{6}-Ga-AlN& -1.321& 0& -1.823 \\
Al\textsubscript{5}GaN\textsubscript{6}-GaN-Ga& -2.030& 0& -1.232 \\
Al\textsubscript{5}GaN\textsubscript{6}-N\textsubscript{2}-AlN& -3.144& -1.823& 0 \\
Al\textsubscript{5}GaN\textsubscript{6}-GaN-N\textsubscript{2}& -3.263& -1.232& 0 \\
\bottomrule
\end{tabular}
\caption{Three phase equilibrium regions of Al\textsubscript{5}GaN\textsubscript{6} in the ternary Al-Ga-N chemical space. $\Delta\mu_{\mathrm{i}}$ (i=Al,Ga,N) are the deviations in the elemental chemical potential from the reference values.}
\end{table}
\begin{table}[h!]
\centering
\renewcommand{\arraystretch}{1.2}
\begin{tabular}{cccc}
\toprule
Limit & $\Delta\mu_{\mathrm{Al}}$ & $\Delta\mu_{\mathrm{Ga}}$ & $\Delta\mu_{\mathrm{N}}$ \\
\midrule
Al\textsubscript{3}GaN\textsubscript{4}-Ga-AlN& -1.439 & 0 & -2.067  \\
Al\textsubscript{3}GaN\textsubscript{4}-GaN-Ga& -2.070 & 0 & -1.593  \\
Al\textsubscript{3}GaN\textsubscript{4}-N\textsubscript{2}-AlN& -3.505 & -2.067 & 0  \\
Al\textsubscript{3}GaN\textsubscript{4}-GaN-N\textsubscript{2}& -3.663 & -1.593 & 0  \\
\bottomrule
\end{tabular}
\caption{Three phase equilibrium regions of Al\textsubscript{3}GaN\textsubscript{4} in the ternary Al-Ga-N chemical space. $\Delta\mu_{\mathrm{i}}$ (i=Al,Ga,N) are the deviations in the elemental chemical potential from the reference values.}
\end{table}
\begin{table}[h!]
\centering
\renewcommand{\arraystretch}{1.2}
\begin{tabular}{cccc}
\toprule
Limit & $\Delta\mu_{\mathrm{Al}}$ & $\Delta\mu_{\mathrm{Ga}}$ & $\Delta\mu_{\mathrm{N}}$ \\
\midrule
Al\textsubscript{2}GaN\textsubscript{3}-Ga-AlN& -1.494 & 0 & -1.650  \\
Al\textsubscript{2}GaN\textsubscript{3}-GaN-Ga& -2.121 & 0 & -1.232  \\
Al\textsubscript{2}GaN\textsubscript{3}-N\textsubscript{2}-AlN& -3.144 & -1.650 & 0  \\
Al\textsubscript{2}GaN\textsubscript{3}-GaN-N\textsubscript{2}& -3.353 & -1.232 & 0  \\
\bottomrule
\end{tabular}
\caption{Three phase equilibrium regions of Al\textsubscript{2}GaN\textsubscript{3} in the ternary Al-Ga-N chemical space. $\Delta\mu_{\mathrm{i}}$ (i=Al,Ga,N) are the deviations in the elemental chemical potential from the reference values.}
\end{table}

\clearpage

\section{S4 Self-consistent Fermi level positions}
\begin{figure}[htbp]
    \centering
    \includegraphics[width=1\textwidth]{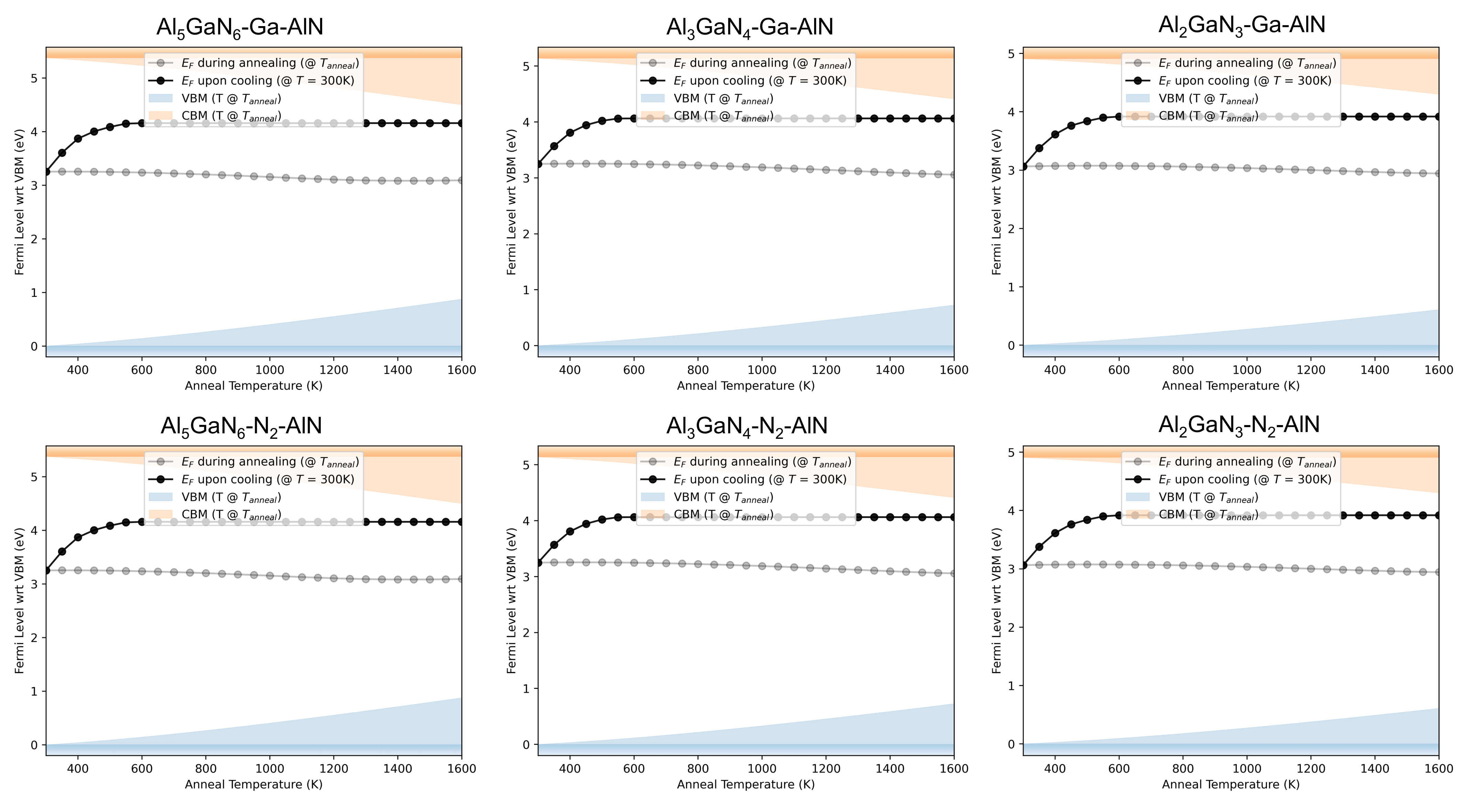}
    \caption{Calculated self-consistent Fermi level positions in Al\textsubscript{1-$x$}Ga\textsubscript{$x$}N during annealing (gray) and upon cooling to room temperature (300 K; black).}
    \label{figure:S3}
\end{figure}

\clearpage
\section{S5 Si-doping in Al$_{1-x}$Ga$_x$N}

        Under N-rich conditions, the deep defect concentrations are $1.48 \times 10^{16}$ ($3.99 \times 10^{18}$), $1.72 \times 10^{16}$ ($1.10 \times 10^{18}$) and $3.71 \times 10^{16}$ ($1.14 \times 10^{18}$) cm$^{-3}$ for x=1/6, 1/4 and 1/3, respectively, without (with) applying the temperature-dependent corrections.
\begin{figure}[htbp]
    \centering
    \includegraphics[width=1\textwidth]{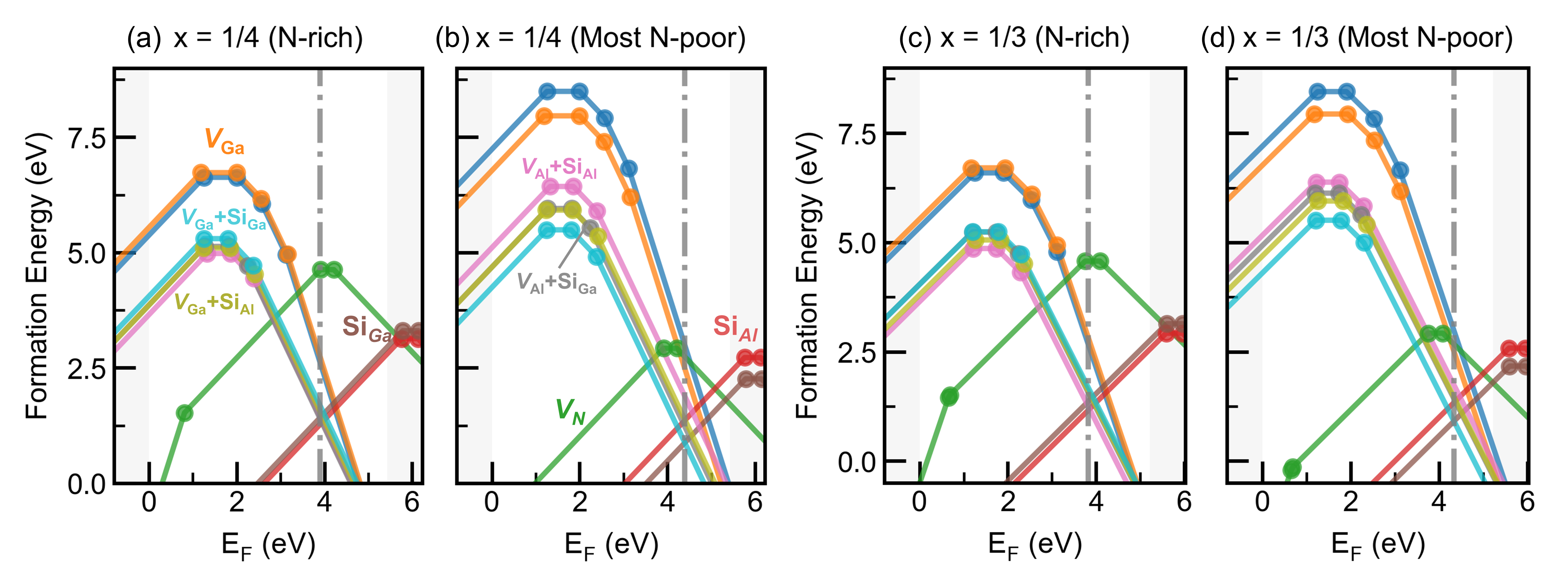}
    \caption{Calculated defect formation energy as a function of the Fermi energy in Si-doped (a,b) Al$_3$GaN$_4$ and  (c,d) Al$_2$GaN$_3$ alloys under both N-rich and most N-poor growth conditions. The equilibrium Fermi energy is calculated at T = 1400 K.}
    \label{figure:S4}
\end{figure}
\begin{figure}[htbp]
    \centering
    \includegraphics[width=0.8\textwidth]{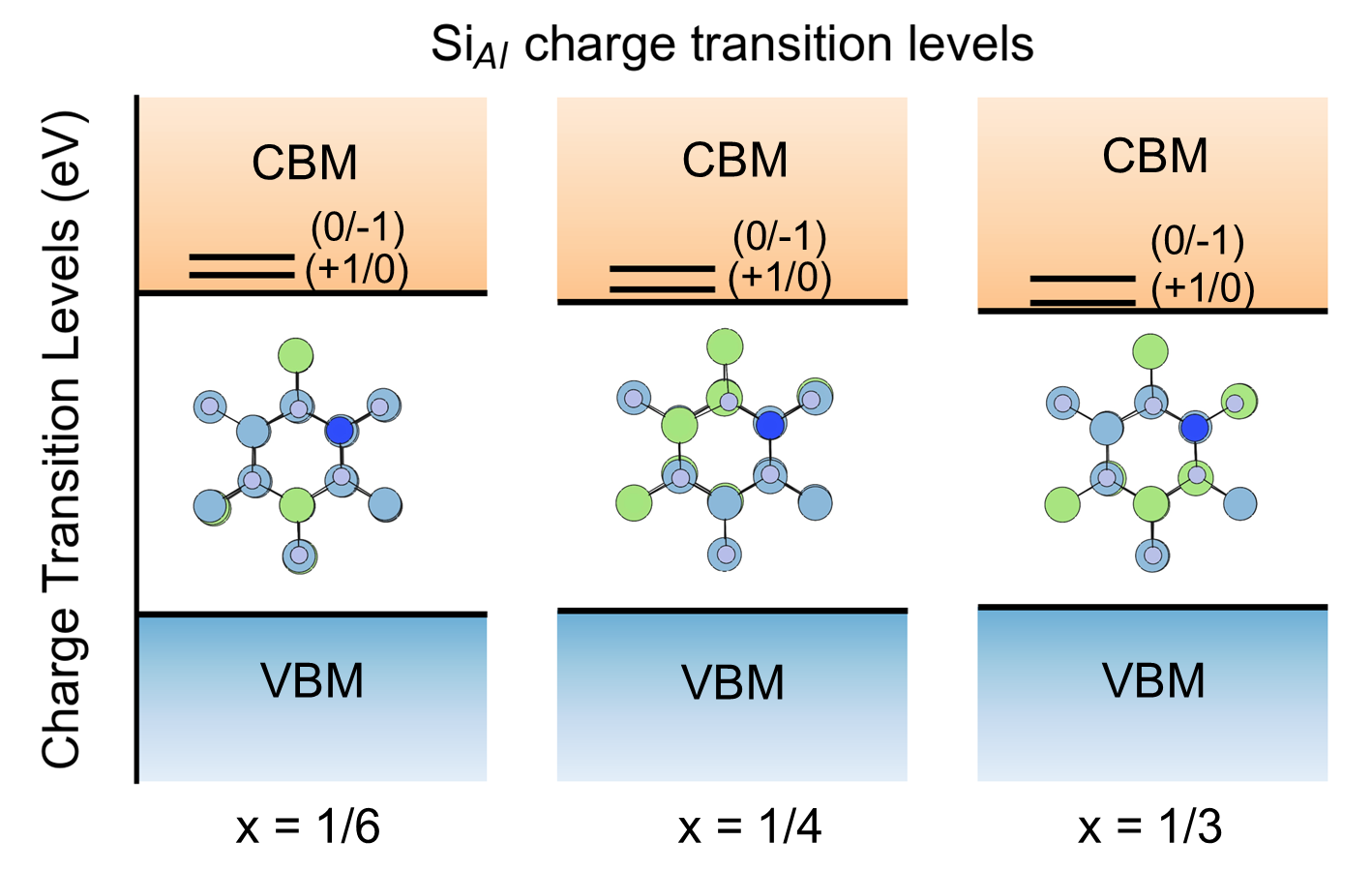}
    \caption{Local structural environments of Si\textsubscript{Al} in Al$_{1-x}$Ga$_x$N alloy. Dark blue, light blue, and green spheres represent Si, Al and Ga atoms, respectively. The Si\textsubscript{Al} site is coordinated by five Ga atoms at the next-nearest-neighbor positions, in contrast to only two Ga atoms for the Si\textsubscript{Ga} site, indicating that the Si\textsubscript{Al} site is embedded in a relatively Ga-rich environment.}
    \label{figure:S5}
\end{figure}
\begin{figure}[htbp]
    \centering
    \includegraphics[width=0.6\textwidth]{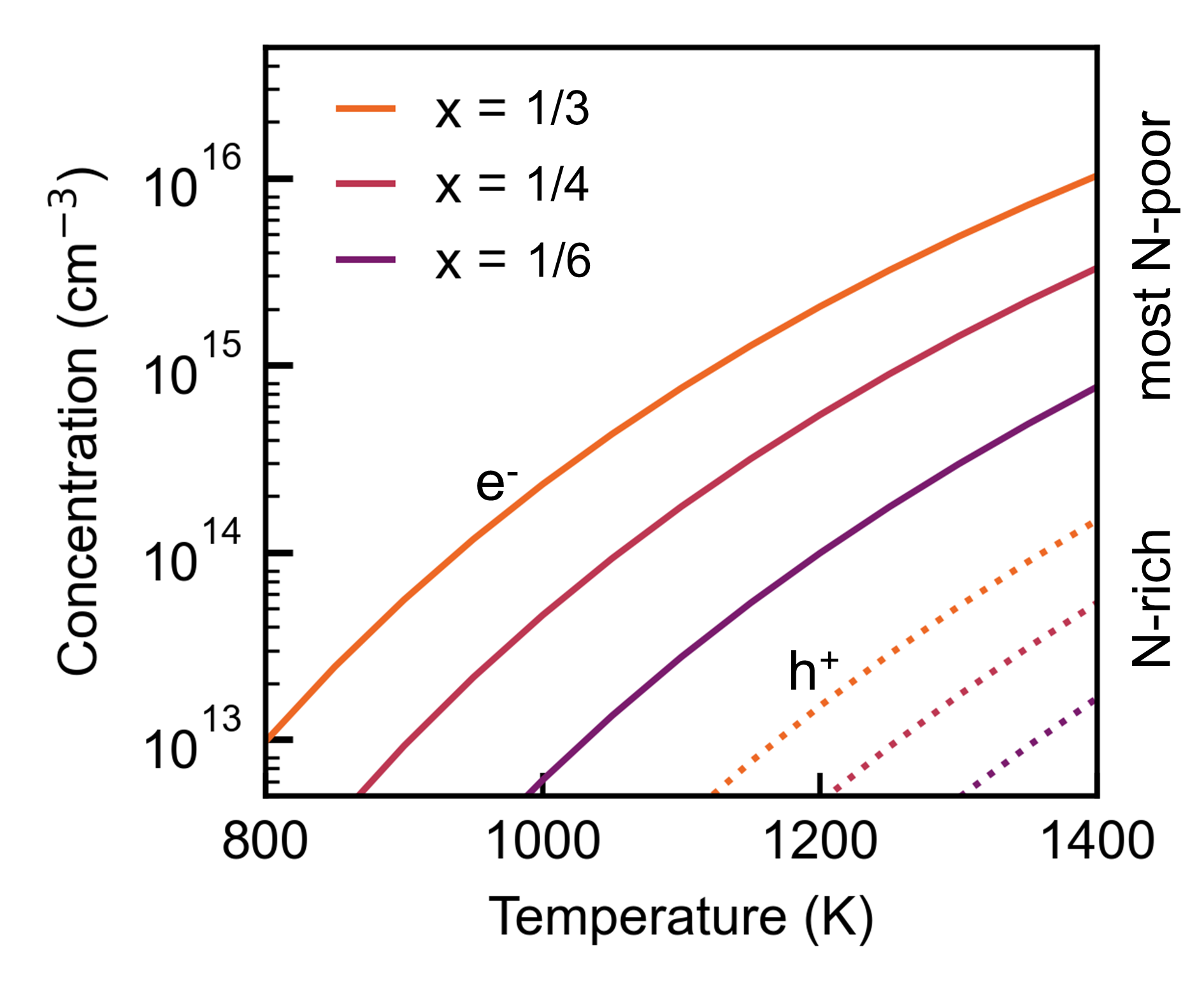}
    \caption{Calculated net carrier concentrations as a function of temperature. The temperature-dependent band gap renormalization are not included.}
    \label{figure:S6}
\end{figure}
\clearpage

\section{S6 Band gap and lattice constant}
\begin{table}[h!]
\centering
\renewcommand{\arraystretch}{1.2}
\begin{tabular}{cccc}
\hline
alpha & bandgap (eV)& \multicolumn{2}{c}{lattice constant (\AA)}     \\ \hline
      &              & $a=b$ & $c$ \\ \hline
0.18  & 5.724        & 3.110 & 4.982 \\
0.19  & 5.819        & 3.109 & 4.981 \\
0.20  & 5.913        & 3.108 & 4.979 \\
0.21  & 6.008        & 3.107 & 4.977 \\
0.22  & 6.100        & 3.107 & 4.973 \\
0.23  & 6.195        & 3.106 & 4.972 \\
0.24  & 6.286        & 3.105 & 4.970 \\
0.25  & 6.381        & 3.104 & 4.968 \\
0.26  & 6.476        & 3.103 & 4.967 \\ \hline
\end{tabular}
\caption{Bandgap and lattice constant as a function of exact exchange fraction $\alpha$. }
\end{table}

\clearpage
\section{S7 Dielectric tensor of AlN and Al\textsubscript{1-x}Ga\textsubscript{x}N }
\begin{table}[htbp]
\centering
\setlength{\tabcolsep}{24pt}
\begin{tabular}{lccc}
\hline
x & $\varepsilon_{xx}$ & $\varepsilon_{yy}$ & $\varepsilon_{zz}$ \\
\hline
0              & 7.15 & 7.15 & 8.56 \\
1/6            & 7.33 & 7.33 & 8.63 \\
1/4            & 7.44 & 7.45 & 8.70 \\
1/3            & 7.56 & 7.55 & 8.77 \\
\hline
\end{tabular}
\caption{Calculated total dielectric tensor of AlN and Al\textsubscript{1-x}Ga\textsubscript{x}N alloys.}
\end{table}
\clearpage